\documentclass[onecolumn,secnumarabic,amssymb, amsmath, nofootinbib,tightenlines,nobibnotes, aps, prf]{revtex4-1}
\usepackage{longtable}
\usepackage{bm}
\expandafter\ifx\csname package@font\endcsname\relax\else
 \expandafter\expandafter
 \expandafter\usepackage
 \expandafter\expandafter
 \expandafter{\csname package@font\endcsname}%
\fi
\usepackage{overpic}
\usepackage{tikz}
\newcommand{\beq}{\begin{equation}}
\newcommand{\eeq}{\end{equation}}
\newcommand{\lb}{\left(}
\newcommand{\rb}{\right)}
\newcommand{\un}[1]{\,\mathrm{#1}}
\usepackage{url}
\usepackage{pgfplots}

\begin{document}

\title{Hysteretic wave drag in shallow water}

\author{GP Benham}
\email{graham.benham@ladhyx.polytechnique.fr}
\affiliation{LadHyX, UMR CNRS 7646, Ecole polytechnique, 91128 Palaiseau, France}
\author{R. Bendimerad}
\affiliation{LadHyX, UMR CNRS 7646, Ecole polytechnique, 91128 Palaiseau, France}
\author{M. Benzaquen}
\affiliation{LadHyX, UMR CNRS 7646, Ecole polytechnique, 91128 Palaiseau, France}
\author{C Clanet}
\affiliation{LadHyX, UMR CNRS 7646, Ecole polytechnique, 91128 Palaiseau, France}

\date{\today}%

\begin{abstract}
During motion from deep to shallow water, multiple equilibria may emerge, each with identical drag - a phenomenon that can be explained by a localised amplification of the wave drag near the shallow wave speed. 
The implication of this is the emergence of several previously unstudied bifurcation patterns and hysteresis routes. 
Here, we address these nonlinear dynamics by considering the quasi-steady motion of a body between deep and shallow water, where the depth is slowly varying. 
We survey several theoretical models for the drag, compare these against our tow-tank experimental measurements, and then use the validated theory to explore the bifurcation patterns using two parameters: the depth of motion and the forcing.
In particular, using a case study of a lake with a sinusoidal depth profile, we illustrate that hysteresis effects can play a significant role on the speed of motion and journey time, presenting interesting implications for naval and racing applications.
\end{abstract}

\maketitle

\section{Introduction}

The effect of bathymetry, or water depth, on the drag of an object moving through a body of water has numerous important applications, ranging from naval and coastal engineering to boat sports and swimming \citep{boucher2018thin,fourdrinoy2019naval}. It is well known that there is significant difference between the waves generated by a body moving in deep water and shallow water, and this is reflected in the drag properties. For example, as first discussed by \citet{havelock1922effect}, wave drag may be either exacerbated or diminished in shallow water, depending on the speed regime. 
However, the catastrophic emergence of multiple equilibrium states in shallow water and the resultant nonlinear dynamics have not yet been discussed. 
{In this study, we investigate these dynamics, including possible bifurcation and hysteresis behaviour, in the case of body motion in water with slowly varying depth. }

To distinguish between deep and shallow water regimes, a commonly used parameter is the non-dimensional ratio between the water depth $h$ and the typical wave length $\lambda$. The dispersive properties of deep water ($h/\lambda\gg 1$) mean that the wave drag behaviour is largely dominated by a resonance between the boat speed $u$ and the wave speed $c$ \citep{havelock1922effect}. The boat speed is usually characterised by the length-based Froude number $\mathrm{Fr}_L=u/\sqrt{gL}$, where $L$ is the boat length. Wave drag is typically largest within a vicinity of $\mathrm{Fr}_L\approx 0.5$, and its importance decays as $\mathrm{Fr}_L\rightarrow 0,\infty$ \citep{michell1898xi,videler2012fish}. By contrast, the non-dispersive properties of shallow water ($h/\lambda\ll 1$) mean that no such resonance exists. Instead, since all waves travel at the same speed $\sqrt{gh}$, the only resonance is isolated to the vicinity where the depth-based Froude number is near unity $\mathrm{Fr}_h=u/\sqrt{gh}\sim 1$. The collapse of this resonance has been compared to shock behaviour in hydraulic jumps, and supersonic transition in aeronautics \citep{tuck1989wave}.

The consequence is that when a boat moves from deep water to shallow water, and vice versa, there is an interesting and unintuitive change of behaviour, as first discussed by \citet{havelock1922effect} and later by various other authors \citep{tuck1989wave,hofman2000shallow}. There are three definitive regimes of interest: subcritical, critical and super-critical. The sub-critical region $\mathrm{Fr}_h\ll 1$ corresponds to situations where the depth is sufficiently large that it makes little impact on the drag. The critical regime $\mathrm{Fr}_h\sim 1$ exhibits resonance with the shallow wave speed, causing a drastic accentuation of the wave drag. By contrast, in the very shallow super-critical regime $\mathrm{Fr}_h\gg1$, the wave drag is less than it would be in infinitely deep water. Hence, as a body is pushed at constant force between deep and shallow water, it may undergo either an acceleration or a deceleration, depending on its speed.

There have been numerous studies of the effects of the bathymetry on the wave drag. One of the most notable is the linear theory of \citet{havelock1922effect}, in which the wave resistance was calculated for a moving pressure disturbance at the surface of a body of water with finite depth. 
Another linear approach, derived by \citet{sretensky1937theoretical} and later summarised in English by \citet{weinblum1950analysis}, adjusted the infinite depth theory of \citet{michell1898xi} to account for both finite depth and width. 
Other studies have included nonlinear effects \citep{doctors1980comparison}, the effects of unsteadiness \citep{doctors1975experimental}, shear currents \citep{li2016ship} and capillary waves \citep{wkedolowski2013capillary}.
{However, whilst several previous experiments have demonstrated that the same drag can be achieved at several different boat velocities (and hence the existence of multiple equilibrium states) \cite{millward1986effect,hofman2006prediction}, the relevant bifurcation behaviour and nonlinear dynamics, including hysteretic motion, have not yet been discussed.

The key to the existence of multiple equilibrium states is a relationship between the drag force and boat velocity which is non-monotone, as is the case for motion in shallow water. 
However, it should be noted that this phenomenon is not unique to shallow water situations. 
A similar non-monotone drag-velocity relationship has also been observed when a boat moves in water with a sharp stable stratification, also known as \textit{dead water}, and possible hysteretic behaviour has even been discussed \cite{ekman1904dead,mercier2011resurrecting,esmaeilpour2018computational,medjdoub2020laboratory}. 
The same phenomenon can also occur due to the interference between bow and stern waves \cite{tuck1987wave}. 
Indeed, some boats are designed in such a way to amplify this effect, thereby modifying their drag curve to place operational speeds near a local minimum in the drag curve, as is the case with the design of the classic bulbous bow \cite{kracht1978design}.
However, as mentioned above, the bifurcation of equilibrium states, and possible hysteresis routes during motion between deep and shallow water have not yet been treated.
In particular, the different solution branches that emerge in shallow water have important consequences on boat racing and naval applications, since they correspond to vastly different speeds. }

In this study we consider the nonlinear dynamics of a boat pushed at constant force at the surface of a body of water with varying depth. We restrict our attention to the case where the length scale associated with depth changes is much larger than the length scale associated with acceleration/deceleration, similarly to \citet{gourlay2003ship}. Such an approximation is realistic in a large number of practical cases. In this way, it is acceptable to consider the wave field steady at each instantaneous depth, and we focus on the quasi-steady dynamics of the boat (where the only important time-dependence is that associated with depth change). 

{
We compare several different theoretical descriptions of the wave drag, including the linear theories of \citet{sretensky1937theoretical} and \citet{havelock1922effect}, as well as the recent boundary layer model of \citet{benham2019wave}. For comparison with these theoretical results, we measure the steady drag at various different speeds and depths using a tow-tank experiment, similarly to previous authors \cite{millward1986effect,hofman2006prediction}.
The main objective of our experimental work is to test our theoretical predictions for the hull shape we use and the relevant range of flow conditions. In this way, we can confidently use these theoretical predictions to investigate the bifurcation behaviour and nonlinear dynamics, which are the focus of this study.
As a further check in our comparison, we also include the experiments of the European Development Centre for Inland and Coastal Navigation, Duisburg (also known as the VBD reports) \cite{hofman2006prediction}.
After surveying these different results, we employ our validated theory to explore the bifurcation patterns of the system, using the two relevant bifurcation parameters: the forcing and the depth. }Then, we describe the quasi-steady motion of a boat in a lake with a sinusoidal depth profile as a case study, revealing interesting speed dynamics and possible hysteresis routes.

\section{Motion in deep and shallow water\label{secdeep}}

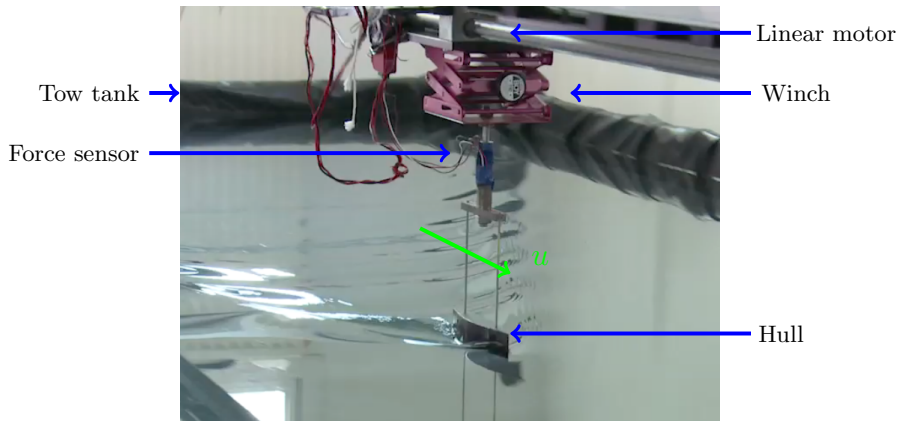
\begin{figure}
\centering
\begin{tikzpicture}[scale=0.4]
\node at (0,0) {\includegraphics[width=0.4\textwidth]{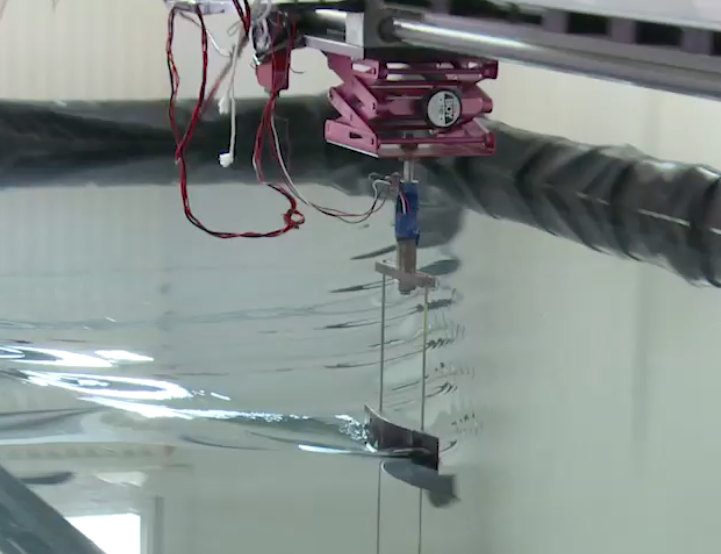}};
\draw[blue,line width=1.5,->] (10,6) -- (2,6);
\draw[blue,line width=1.5,->] (10,4) -- (4,4);
\draw[blue,line width=1.5,->] (10,-4) -- (2,-4);
\draw[blue,line width=1.5,->] (-10,2) -- (0,2);
\draw[blue,line width=1.5,->] (-10,4) -- (-9,4);
\node at (-12.5,2) {Force sensor};
\node at (-12,4) {Tow tank};
\node at (12.5,6) {Linear motor};
\node at (11.5,4) {Winch};
\node at (11,-4) {Hull};
\draw[green,line width=1.5,->] (-1,-0.5) -- (2,-2);
\node[green] at (3,-1.5) {\large  $u$};
\end{tikzpicture}
\caption{Labelled photograph of our experimental setup: a hull is pulled through the water at constant velocity by a linear motor and connected via supporting bars to a force sensor. The water depth for each experiment is uniform. \label{schem}}
\end{figure}

We consider the motion of a body of mass $m$ moving in a large region of water with finite depth. Written in terms of the Cartesian coordinate system ($x,y,z$), the body motion is purely in the $x$-direction. The $z$-axis is defined such that the resting water surface is given by $z=0$, and the water subsurface only varies in the direction of motion $z=-h(x)$. 
The body is driven forwards with a constant force $F$, and the position and speed $x(t),u(t)$ are governed by the dynamics
\begin{align}
    \dot{x}&=u,\label{sys1}\\
    (m+m_a)\dot{u}&=F-R_d(x,u),\label{sys2}
\end{align}
where $R_d$ is the drag force, which depends on both space (via the water depth) and the boat speed, and $m_a$ is the added mass \citep{newman2018marine}.
We make the key assumption that the depth is sufficiently slowly varying that any acceleration/deceleration due to depth change is small, or equivalently
\beq
\dot{u}/g\ll 1\label{smallacc}.
\eeq
{Body acceleration is compared to gravitational acceleration in (\ref{smallacc}) because this is the dominant acceleration scale associated with surface gravity waves}.
Under the assumption (\ref{smallacc}), the drag term in \eqref{sys2} is approximated as its steady value at each instantaneous depth and speed (i.e. ignoring any dependence on acceleration).

As is often done, we decompose the drag into three different components: the wave drag, form drag and skin drag, such that
\beq
R_d=R_w+R_f+R_s.\label{dragadd}
\eeq
The wave drag $R_w$ is the force associated with the sustained generation of waves due to body motion, and depends greatly on both the speed and the draft \citep{benham2019wave}, as well as the depth of the water \citep{havelock1922effect}. 
Typical speed values for different types of boats and aquatic animals are given by \citet{boucher2018thin}, 
and typical depth ratios are anything larger than around $h/L=0.2$.
The form drag $R_f$ is the force due to the surrounding pressure field and is effectively a measure of how streamlined an object is. Finally, the the skin drag $R_s$ relates to the viscous  friction between the wetted body surface and the surrounding fluid. 

{In the next step, we present experimental drag measurements for a symmetric hull across a range of depths and speeds. As discussed earlier, the main purpose of these experiments is as a means of testing the validity of our theoretical predictions for the parameters and hull shape we use, so we can reliably and accurately use them to study the bifurcation behaviour and nonlinear dynamics in the subsequent sections.}
We measure the total drag force $R_d$ on a 3D-printed hull experimentally by pulling it through a large basin of water at constant velocity using a linear motor (see Fig. \ref{schem}). 
{We use a symmetric hull, whose shape varies only in the direction of motion $x$, not in the vertical $z$, and is hence represented by a function $y/L=\hat{f}(\hat{x})$, where $\hat{x}=x/L$, and}
\beq
{\hat{f}(\hat{x})=c_1\log \lb \frac{1+c_2}{e^{c_3 (\hat{x}-1/2)}+c_2 e^{-c_3 (\hat{x}-1/2)}} \rb,}\label{shapefun}
\eeq
and the coefficients in \eqref{shapefun} are given by $c_1=0.0767,\,c_2=0.0302,\,c_3=3.5$.
The dimensions of the submerged\footnote{Note that the total hull height is $5\un{cm}$, but only half of it is submerged beneath the water.} hull in the $(x,y,z)$ directions are $(L,W,H)=(18,3,2.5)\un{cm}$, giving aspect ratios in the horizontal and vertical directions as $\alpha=L/W=6$ and $\beta=L/H=7.2$, respectively.

The hull is connected to the motor by two supporting bars and a force/displacement sensor. 
{The force sensor consists of a set of strain gauges which measure the deformation of the supporting bar (where the linear force-deformation relationship is determined by a calibration step)}. We study several different values of the water depth, $h/L=\{1.11,0.81,0.5,0.33,0.24\}$, and Fr$_L$ values between 0.2 and 1, and in each case the water depth is uniform throughout the basin. 
{We do not study outside this range of Fr$_L$, since the signal to noise ratio for force measurements becomes too large.}
In all cases, the experiments are repeated at least 3 times for accuracy.
We do not allow the pitch of the boat to change, and we do not study the effects of planing \citep{rabaud2014narrow}. 
These measurements for uniform depth will be used later to approximate the instantaneous drag in a medium with slowly varying depth (under the quasi-steady assumption).

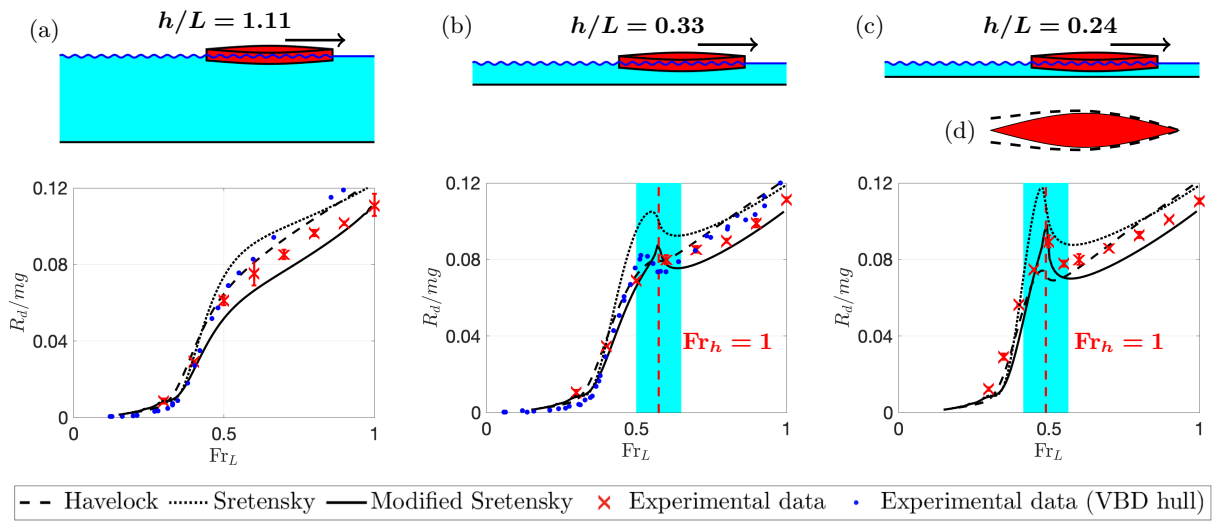
\begin{figure}
\centering
\begin{minipage}{0.3\textwidth}
\begin{tikzpicture}[scale=0.4]
\node at (5.5,9) {\includegraphics[width=1\textwidth]{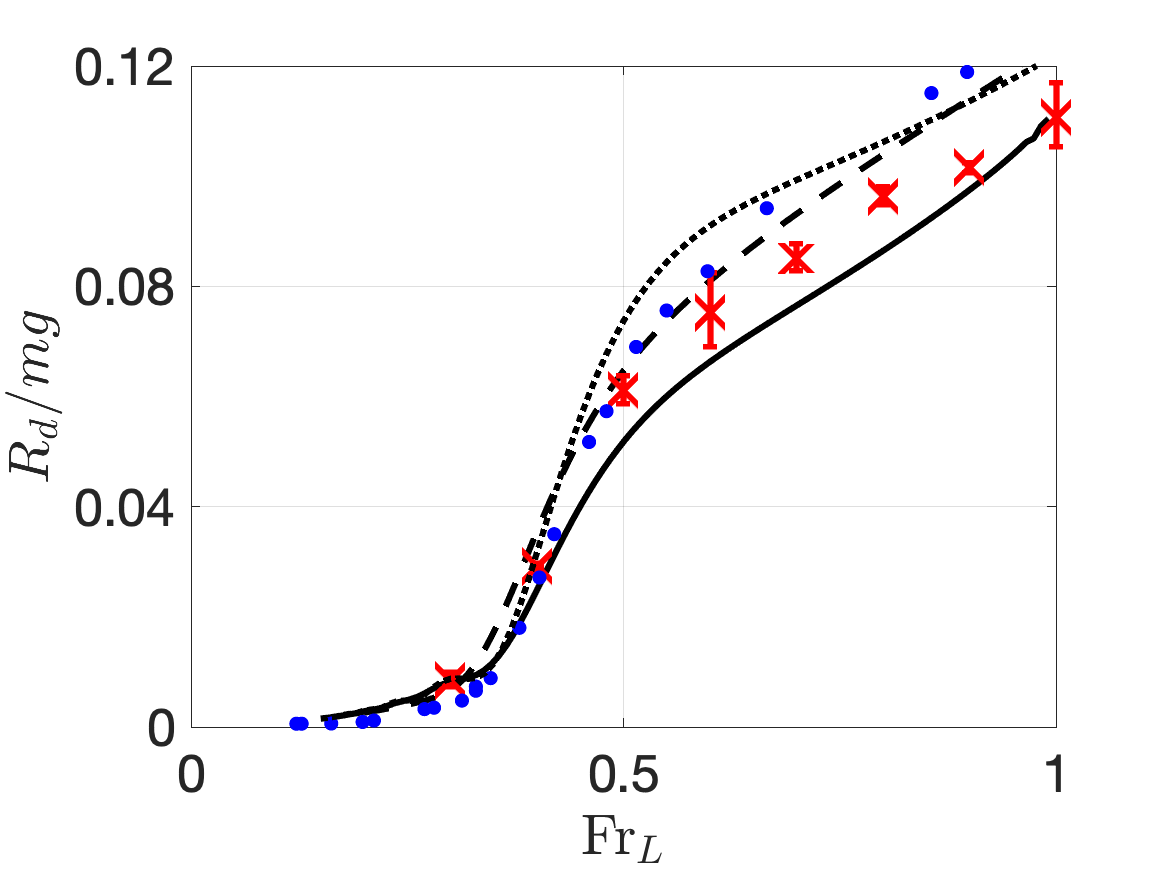}};
\node at (5.5,16) {\includegraphics[width=1\textwidth]{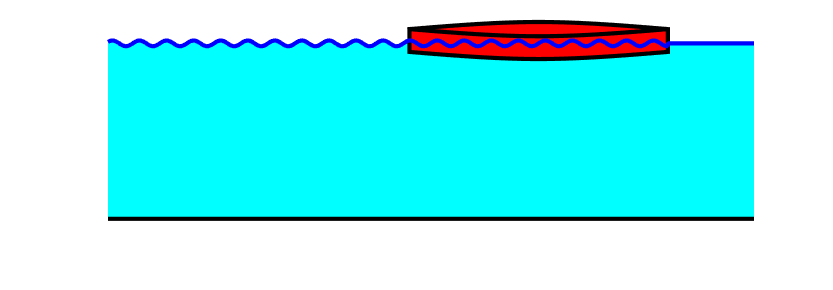}};
\draw[line width=1,->] (8,18.2) -- (10,18.2);
\node at (6,18.8) { $\boldsymbol{h/L=1.11}$};
\node at (0,18.5) { (a)};
\end{tikzpicture}
\end{minipage}
\begin{minipage}{0.3\textwidth}
\begin{tikzpicture}[scale=0.4]
\node at (5.5,9) {\includegraphics[width=1\textwidth]{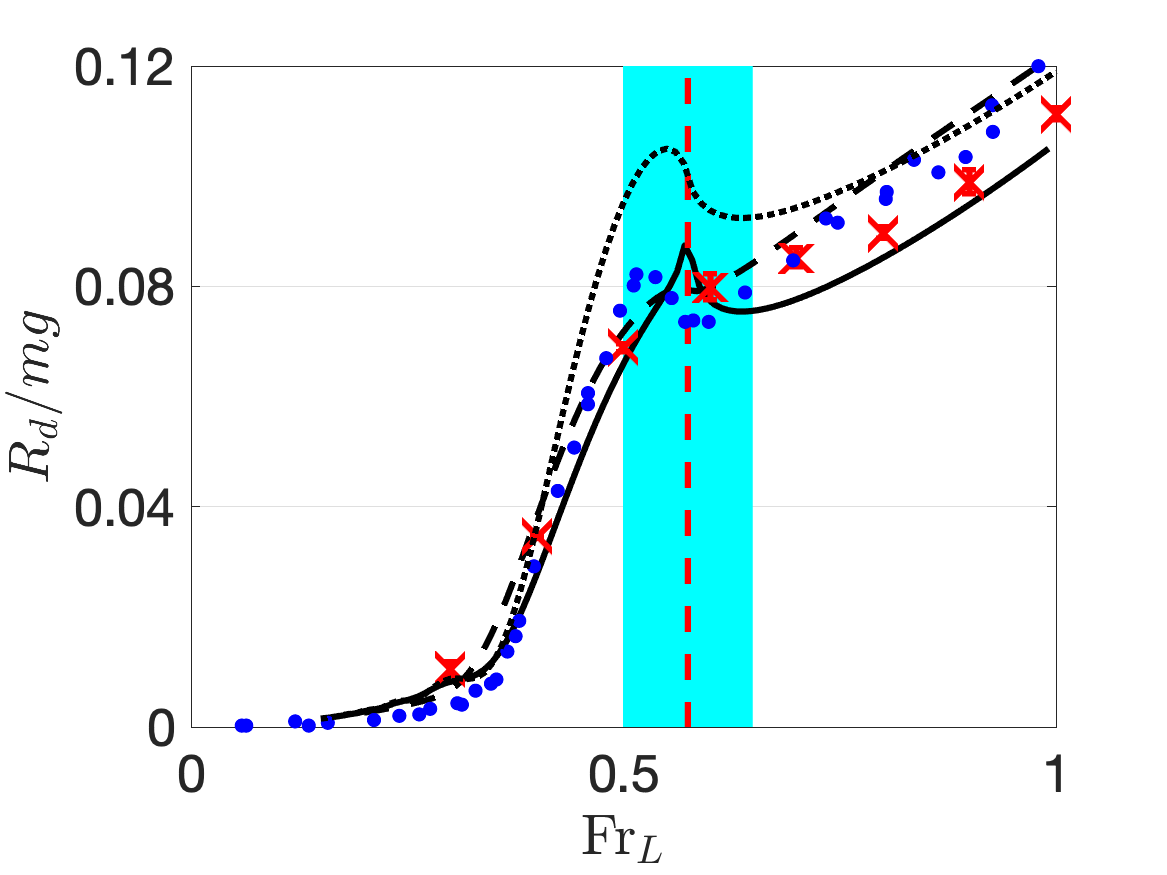}};
\node at (5.5,16) {\includegraphics[width=1\textwidth]{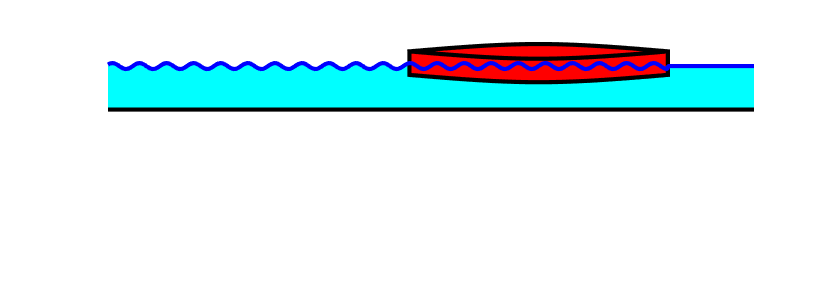}};
\draw[line width=1,->] (8,17.9) -- (10,17.9);
\node at (6,18.5) { $\boldsymbol{h/L=0.33}$};
\node[red] at (9,8) { \bf Fr$\boldsymbol{_h=1}$};
\node at (0,18.5) { (b)};
\end{tikzpicture}
\end{minipage}
\begin{minipage}{0.3\textwidth}\begin{tikzpicture}[scale=0.4]
\node at (5.5,9) {\includegraphics[width=1\textwidth]{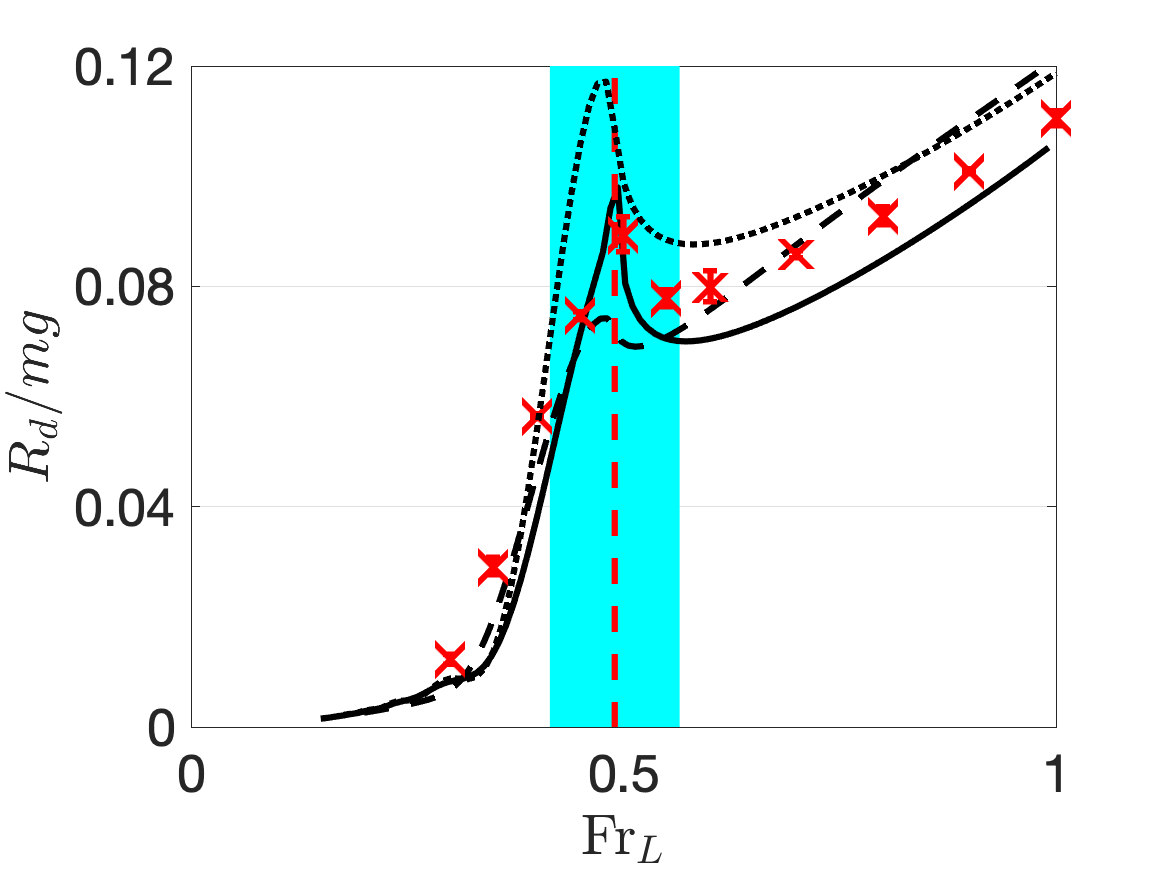}};
\node at (5.5,16) {\includegraphics[width=1\textwidth]{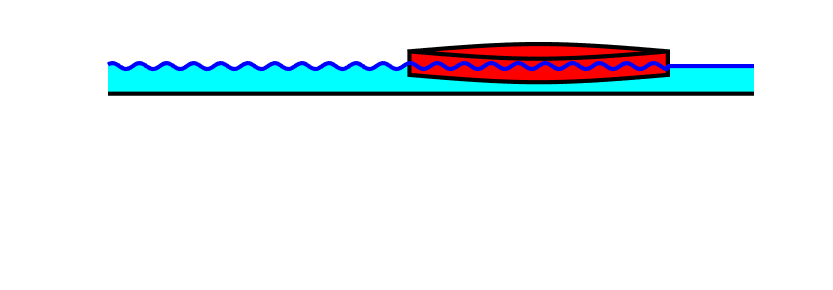}};
\draw[line width=1,->] (8,17.9) -- (10,17.9);
\node at (6,18.5) { $\boldsymbol{h/L=0.24}$};
\node[red] at (8.3,8) { \bf Fr$\boldsymbol{_h=1}$};
\node at (0,18.5) { (c)};
\node at (7,15) {\includegraphics[width=0.6\textwidth]{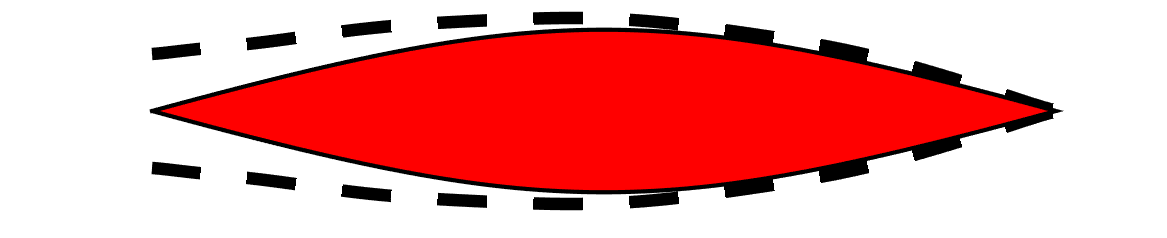}};
\node at (3,15) { (d)};
\end{tikzpicture}
\end{minipage}
\\
\includegraphics[width=0.9\textwidth]{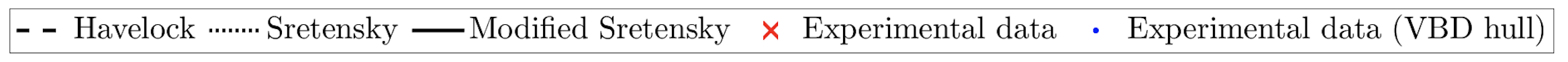}\\
\caption{(a,b,c) Comparison between experimental measurements and theoretical predictions of the drag force for different depths and Froude numbers, and an illustration of each depth case above. 
{Theoretical $R_w$ predictions are summed with wind tunnel measurements of $R_f+R_s$ for comparison with tow-tank measurements of $R_d$ (see discussion in the main text).}
The critical depth-based Froude number Fr$_h=u/\sqrt{g h}=1$ is indicated in each plot, and areas of accentuated drag are illustrated with shading. An illustration of the calculated boundary layer profile for our hull shape, which we use in our modification to Sretensky's theory, is given in (d) {(see discussion towards the end of Section \ref{secdeep})}. 
\label{expresults}}
\end{figure}

In Fig. \ref{expresults} we plot the drag force $R_d$ for various different values of the water depth $h/L$ and the length-based Froude number $\mathrm{Fr}_L$ (though we omit the cases $h/L=0.5,0.81$ since they have effectively no depth effects, and are very similar to $h/L=1.11$). In each case, we plot the drag in terms of the typical weight needed to balance the Archimedes buoyancy force of a boat of such volume $mg=\rho gL^3/\alpha\beta$.
Error bars correspond to one
standard deviation of the time signal given by the force sensor.
In addition to our drag measurements, we also plot the drag measurements taken from the VBD reports, as detailed by \citet{hofman2006prediction}, for which a Taylor Standard series ship hull shape is used (given in terms of a fifth order polynomial \citep{gertler1954reanalysis}). For the latter case, we only display the data in the cases where the same depth ratio was available. There are also several theoretical results overlaid on the same plot, which will be discussed shortly.

For depths larger than $h/L\approx 0.4$, the drag behaves similarly to the case of infinite depth, with a monotone increasing curve. However, for depths smaller than this, resonance with the shallow water wave speed accentuates the drag in an isolated region near Fr$_h=1$, causing a pinching of the curve.
This results in the emergence of a stationary point (maximum) near $\mathrm{Fr}_h=1$, and another stationary point (minimum) for a slightly larger $\mathrm{Fr}_L$. Indeed, there are multiple values of the length-based Froude number that result in the same drag force, {as reported experimentally by other authors \cite{hofman2000shallow,millward1986effect}}. This implies that by pushing a boat with constant force in the critical regime, as many as three different velocities could be achieved. 

The next step is to compare these results to available theoretical predictions.
A full derivation of the theories used here can be found in the original references, though for our purposes we simply state the formulae. Two commonly used formulae for the wave drag are the linear theories of \citet{sretensky1937theoretical} and \citet{havelock1922effect}. Sretensky's formula, which is an extension of Michell's formula for the wave drag at infinite depth \citep{michell1898xi}, assumes an inviscid, irrotational, steady flow past a slender body. The wave drag is given in non-dimensional terms by the formula
\beq
\frac{R_w}{mg}=\frac{2{\alpha}\beta\mathrm{Fr}_L}{\pi}\int_{\hat{k}_0}^\infty |I(\hat{k})|^2\frac{ \lb \sinh \hat{k}\hat{h} - \sinh \hat{k}(\hat{h}-1/\beta)\rb^2}{\hat{k}\cosh^2 \hat{k}\hat{h} \sqrt{\mathrm{Fr}_L^2 \hat{k}^2-\hat{k}\tanh \hat{k}\hat{h} } }  \, \mathrm{d}\hat{k},\label{sretint}
\eeq
where the function $I(\hat{k})$ is given by
\beq
I(\hat{k})=\int_{-1/2}^{1/2} \hat{f}'(\hat{x})  e^{i \hat{x} \sqrt{\hat{k}/\mathrm{Fr}_L^2\tanh \hat{k}\hat{h}}}\,\mathrm{d}\hat{x},
\eeq
and hats denote variables non-dimensionalised with respect to the body length (e.g. $\hat{h}=h/L$). The minimum non-dimensional wave number in the integral \eqref{sretint}, $\hat{k}_0$, is given by the formula
\beq
\mathrm{Fr}_L^2 \hat{k}_0 - \tanh \hat{k}_0\hat{h} =0,
\eeq
{which corresponds to when the body speed matches the wave speed.}
Note that Sretensky's formula is not capable of predicting skin and form drag since it neglects viscosity. However, we can measure these using a simple wind tunnel experiment and add them (as in \eqref{dragadd}) to the theoretical prediction of the wave drag. 
{Over the range of depths and speeds we consider here, the skin and form drag are very well approximated by the formula $(R_f+R_s)/mg=0.017\cdot\mathrm{Fr}_L^2$.}

Unlike Sretensky's formula which is written in terms of the hull shape $\hat{f}(\hat{x})$, the theory of \citet{havelock1922effect} relies on knowledge of the pressure disturbance caused by the hull instead. Since it is difficult to know the pressure disturbance exactly, it is often assumed to be a Gaussian distribution with the appropriate aspect ratio values \cite{benzaquen2014wake}. Hence, we take the surface pressure distribution to be
\beq
p(\hat{x},\hat{y})/P_0 =  e^{-\lb \hat{x}^2/2+\alpha^2\hat{y}^2/2 \rb},\label{havpress}
\eeq
where the magnitude of the pressure distribution $P_0$ is unknown, but should be of the same order of magnitude as the pressure needed to support a surface disturbance of height comparable to the draft. This can be confirmed by our observation of typical wave amplitudes compared to the total draft. Hence, we take $P_0=\kappa \rho g H$, where the choice of $\kappa$ is discussed later. Given this pressure field, Havelock's formula (after simplification) gives the drag as
\beq
\frac{R_w}{mg}=\frac{\kappa^2}{4 \mathrm{Fr}_L^2\alpha\beta}\int_0^\infty \frac{ a(\hat{k}_x^*(\hat{k}_y),\hat{k}_y)}{\partial b/\partial \hat{k}_x(\hat{k}_x^*(\hat{k}_y),\hat{k}_y)}\, \mathrm{d}\hat{k}_y,
\eeq
where the functions $a$ and $b$ are given by
\begin{align}
a(\hat{k}_x,\hat{k}_y)&= e^{-\frac{1}{2\pi^2}(\hat{k}_x^2+\hat{k}_y^2/\alpha^2)}{ \hat{k}\tanh \hat{k}\hat{h}},\\
b(\hat{k}_x,\hat{k}_y)&= (\hat{k} \tanh \hat{k}\hat{h})/({\hat{k}_x}\mathrm{Fr}_L^2)-{\hat{k}_x},
\end{align}
and the wavenumber magnitude is $\hat{k}=\sqrt{\hat{k}_x^2+\hat{k}_y^2}$.
Finally, the critical wavenumber $\hat{k}_x^*(\hat{k}_y)$ is given by
\beq
b(\hat{k}_x^*(\hat{k}_y),\hat{k}_y)=0.
\eeq

A recent study by \citet{benham2019wave} showed that the wave drag at infinite depth can be well-approximated by a modification to the theory of \citet{michell1898xi} (the precursor to that of Sretensky), where the shape $f(\hat{x})$ is replaced by the combined shape of the hull and its surrounding boundary layer profile $f(\hat{x})+\delta(\hat{x})$. In particular, the presence of the boundary layer distinguishes the difference between forward and backward motion for an asymmetric hull, which Michell's formula alone is incapable of predicting. Though we focus on symmetric hulls here, we nevertheless make the same modification to Sretensky's formula \eqref{sretint} by including the boundary layer profile, which we extract from a steady $k$-$\omega$ Shear Stress Transport \citep{menter1992improved} numerical simulation (see \citep{benham2019wave} for more details). Our calculated boundary layer profile is illustrated in Fig. \ref{expresults}(d). This modification appears to give better comparison with experimental results than the original.

Theoretical predictions for the wave drag are compared to the experimental results in Fig. \ref{expresults} (a,b,c). 
Qualitatively, all the theoretical approaches capture the drag behaviour, though with varying degrees of accuracy. We find that the best comparison is with Havelock's formula, where the magnitude of the pressure disturbance \eqref{havpress} is fitted to give $\kappa=0.4$, with mean relative error $8\%$. The next best comparison is our modified version of Sretensky's formula (with mean relative error $9\%$), and finally Sretensky's original formula (with mean relative error $13\%$). However, we note that the best accuracy near the critical regime $\mathrm{Fr}_h\sim 1$ is exhibited by our modified version of Sretensky's formula, since Havelock's formula under-predicts drag here. {This is possibly due to the pressure distribution \eqref{havpress} having an insufficiently steep bow and stern characteristic. Havelock's integral could be improved with a distribution of potential sources and sinks at the leading and trailing edges of the hull, as discussed by \citet{zhu2015farfield}. Nevertheless, our boundary layer modification to Sretensky's theory provides a sufficiently good representation of the drag. }

We note that linear theory can only be expected to perform well when the wave magnitude is small, and indeed considerably smaller than the water depth $A/h\ll 1$. Otherwise, nonlinear effects are expected to have an important role, such as wave steepening, cresting, breaking, and even wetting the subsurface. However, we restrict our experimental study to the range $h/L>0.24$, and the maximum observed wave amplitude we observe has $A/h\approx 0.25$. Therefore, we conclude that nonlinear effects bear little importance here. Furthermore, we ignore capillary effects since they do not present a significant contribution to the drag \citep{benham2019wave}.

Each of the theoretical predictions confirm that multiple solutions exist near the vicinity of $\mathrm{Fr}_h\sim 1$, though the precise shape of the drag curve differs. Since our modified version of Sretensky's formula shows the best comparison with experimental data near the critical regime, we choose this theory to provide a means of predicting the drag in between our experimentally measured points. In this way, we can accurately describe the motion of a boat with constant forcing using the quasi-steady system \eqref{sys1}-\eqref{sys2}.

\section{Bifurcation diagrams and stability criteria}

\begin{figure}
\centering
\begin{tikzpicture}[scale=0.67]
\node at (0,0) {\includegraphics[width=0.45\textwidth]{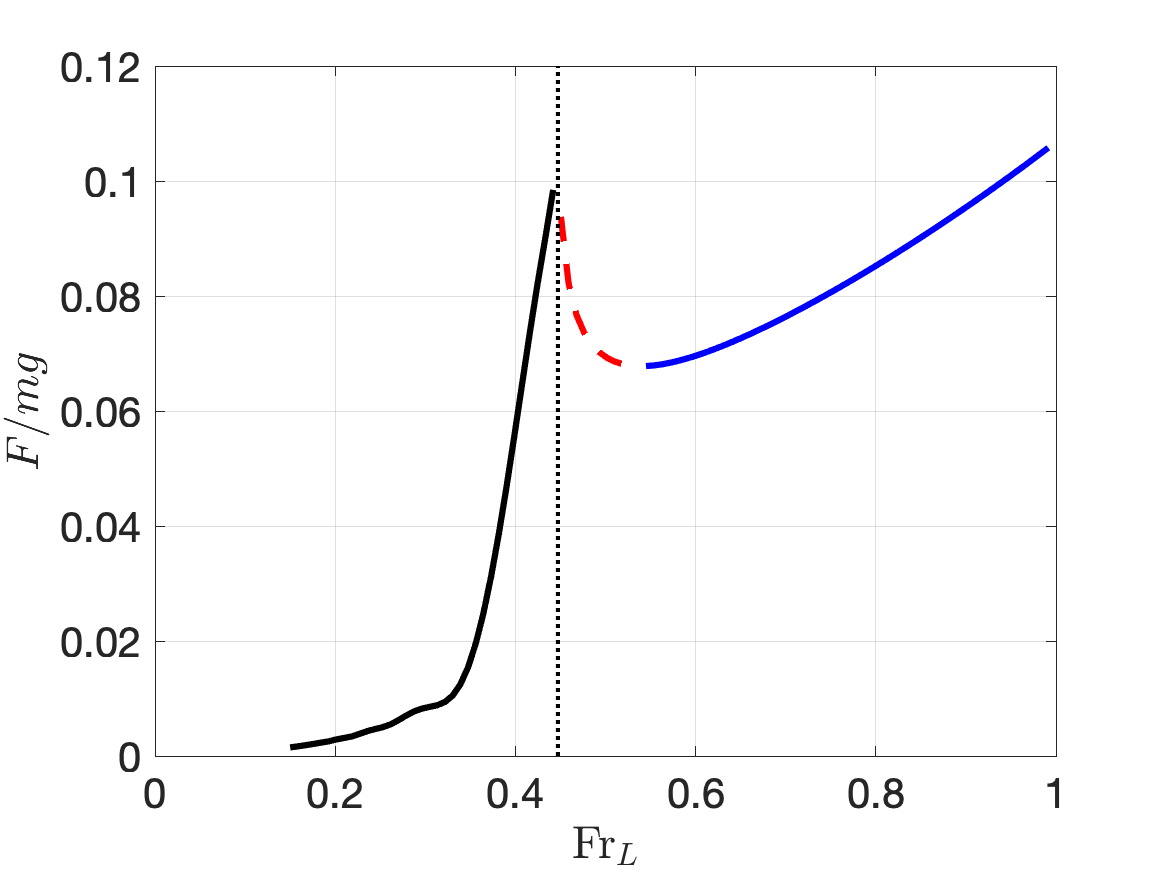}};
\node at (2,0.25){  Fr$_h\gg1$};
\node at (2,-0.25){ \it  Super-critical} ;
\node at (-1.5,2){  Fr$_h\sim1$} ;
\node at (-1.5,1.5){ \it  Critical} ;
\node at (-2.5,-1.5){  Fr$_h\ll1$} ;
\node at (-2.5,-2){ \it  Sub-critical} ;
\draw[line width=2,->,cyan] (-0,2.4) -- (3.8,2.4);
\draw[line width=2,->,cyan] (0.6,0.7) -- (-0.6,0.7);
\draw[line width=0.8,-,black, dotted] (-4.5,0.7) -- (-0.6,0.7);
\draw[line width=0.8,-,black, dotted] (5,0.7) -- (0.6,0.7);
\node at (-3,0.4){  $F=F_c$} ;
\node at (-6,4) {(a)};
\node at (0,4.5) {$\boldsymbol{h/L=0.25}$};
\end{tikzpicture}
\begin{tikzpicture}[scale=0.67]
\node at (0,0) {\includegraphics[width=0.45\textwidth]{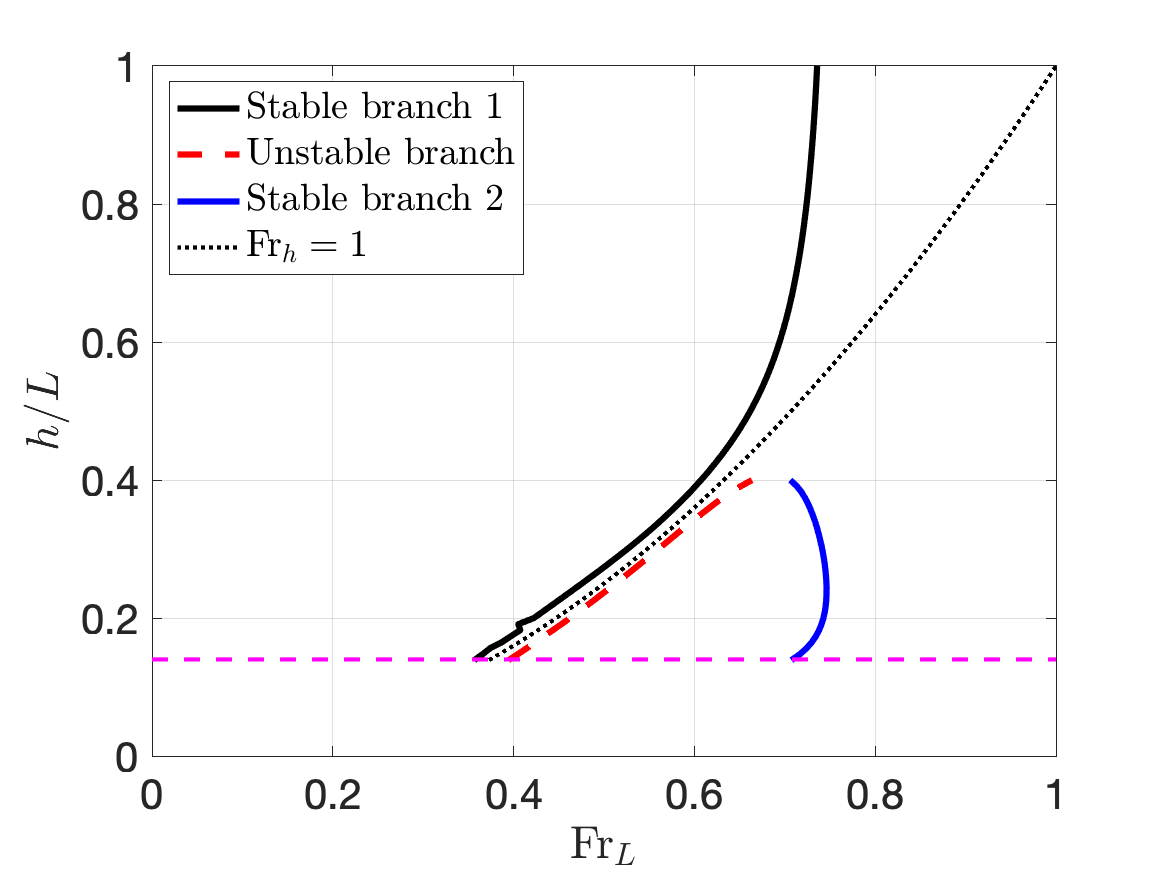}};
\node at (-6,4) {(b)};
\node at (-1,-3) { $h/L\leq 1/\beta$ (no solution)};
\draw[line width=2,-,cyan] (2.8,-2.2) .. controls (2.6,-0.3) and (2.2,-0.2) .. (1.8,-0.2);
\draw[line width=2,->,cyan] (1.8,-0.2) .. controls (2.4,1) .. (2.7,2.5);
\draw[line width=2,->,cyan] (2.1,2.5) .. controls  (1.8,0.5) .. (0.3,-1);
\node at (0,4.5) {$\boldsymbol{F/mg=0.08}$};
\end{tikzpicture}
\caption{Bifurcation diagrams, showing all of the existing branches of the quasi-steady equilibrium solutions to \eqref{sys1}-\eqref{sys2}, for the two bifurcation parameters  $F/mg$ and $h/L$. 
We also illustrate the different sub-critical, critical and super-critical regimes, the critical forcing $F=F_c$ (see discussion in Section \ref{secstudy}), and possible hysteresis routes.  \label{bifurcs}}
\end{figure}
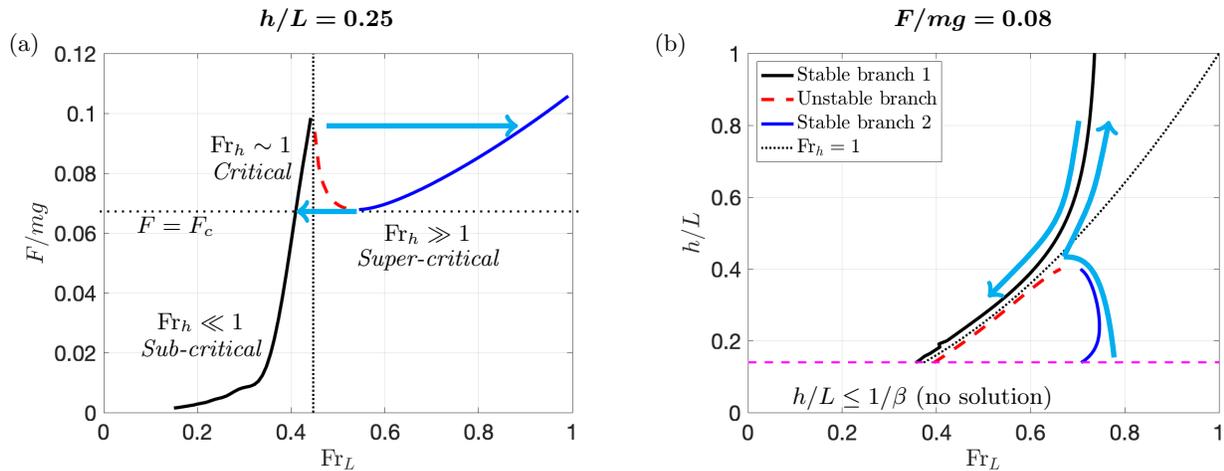

There are two bifurcation parameters which govern the nonlinear dynamics of the system \eqref{sys1}-\eqref{sys2}: the depth $h/L$ and the forcing $F/mg$. 
We display the bifurcation diagrams for each of these parameters in Fig. \ref{bifurcs}, illustrating the different available equilibrium speeds (represented by $\mathrm{Fr}_L$).
To explore the bifurcation parameter $F/mg$, we fix the depth at $h/L=0.25$. In this case, three solutions exists in the range $F/mg\in(0.07,0.095)$. To explore the bifurcation parameter $h/L$, we fix the forcing at $F/mg=0.08$. In this case, three solutions exist in the range $h/L\in(1/\beta,0.4)$ (No solution exists when the depth is shallower than the boat itself $h/L\leq 1/\beta$).

The different solution branches may correspond to significantly different Froude numbers, and hence equilibrium speeds. For example, at forcing $F/mg=0.08$, the available Froude numbers are between 0.4 and 0.7. That is to say, a boat which is pushed at this forcing could be going at three possible speeds, where the fastest speed is almost double the slowest. Evidently, this has significant implications for rowing and sailing races in waters with depth varying between deep and shallow regions, since jumping from one branch to another could result in hugely improved race times, and possibly even new world records.

To assess the stability of these different branches, we now perform a linear analysis. Consider an equilibrium solution $(x_0,u_0)$, such that the right hand side of \eqref{sys1}-\eqref{sys2} is zero. Then consider a small perturbation away from this solution  
\beq
    x=x_0+\epsilon(t),\quad u=u_0+\dot{\epsilon}(t).\label{perturb1}
\eeq
For the simplicity of the stability analysis, we restrict our attention to the case where the depth is constant (such that $\partial R_d/\partial x=0$). Then, inserting \eqref{perturb1} into the system \eqref{sys1}-\eqref{sys2}, expanding in terms of the small variable $\epsilon(t)/L\ll1$, and keeping only leading order terms, we find that
\beq
(m+m_a)\ddot{\epsilon}+\frac{\partial R_d}{\partial u}(x_0,u_0)\dot{\epsilon}=0.
\eeq
In this case, the stability is purely determined by the sign of $\partial R_d/\partial u$, or equivalently $\partial R_d/\partial \mathrm{Fr}_L$. Consequently, all solutions in Fig. \ref{bifurcs}(a) which lie on the part of the drag curve with negative slope are unstable, and all the solutions which lie on the positive slope are stable. Using this knowledge, we have illustrated the corresponding stability of the different branches in the bifurcation diagrams using dashed lines. Likewise, the corresponding unstable branch in Fig. \ref{bifurcs}(b) is also indicated.

We also display several possible hysteresis routes during motion.
For example, in (a) we show how slowly increasing the forcing (starting from zero) would cause a jump increase in velocity at the turning point near Fr$_h=1$. From here, slowly decreasing the forcing, the boat would return on a different solution branch until a jump back to the original branch at the second turning point. Similarly, we show a possible hysteresis route in (b) by varying the depth instead. 
{In this case, 
the depth is slowly increased as the boat moves away from a shallow region to a deep region (initially moving at Fr$_L\approx0.7$ on the fast solution branch). Once the depth reaches a critical value of around $h/L\approx0.4$ (at the fold bifurcation), there is a sudden decrease in velocity as the solution jumps to the slow branch. Upon return to shallow water, the boat remains on this slower stable branch, since the other branches are topologically disconnected.}

\section{It came from the deep...\label{secstudy}}

Having validated the theoretical results against experimental data, and having explored the different available solution branches that occur at various depths and forcing, now we apply the system \eqref{sys1}-\eqref{sys2} to a case study. Using our modified version of Sretensky's theory (for the wave drag), and wind tunnel measurements (for the form and skin drag), we investigate the quasi-steady dynamics of a body moving in a lake of length $nL$ with a sinusoidal depth profile of the form 
\beq
h(x)=1/2\big[(h_{\max}+h_{\min})-(h_{\max}-h_{\min})\cos{2\pi x/nL}\big].\label{sindepth}
\eeq
We choose a large (but realistic) race length $n=400$ to ensure a slowly varying depth, and we set the maximum and minimum depths as $h_{\max}/L=1$ and $h_{\min}/L=0.2$ to ensure we do not venture outside of the region we have studied experimentally (so that our predictions remain accurate). With these values, the maximum calculated value for the acceleration due to depth change is $\dot{u}\approx 0.6\un{m/s^2}$, which is much smaller than $g=9.8\un{m/s^2}$, indicating that the quasi-steady approximation is appropriate.

Motion from shallow to deep water corresponds to $0<x<nL/2$, and likewise from deep to shallow water $nL/2<x<nL$. This canonical depth profile not only exhibits the nonlinear phenomena we have illustrated both experimentally and theoretically, but also is a physically realistic scenario. There are many lakes and rivers that have patches of shallow water, and this study illustrates the possible outcomes that can occur as a boat passes over such a patch.

\begin{figure}
\centering
\begin{tikzpicture}[scale=1.33]
\node at (0,0) {\includegraphics[width=0.45\textwidth]{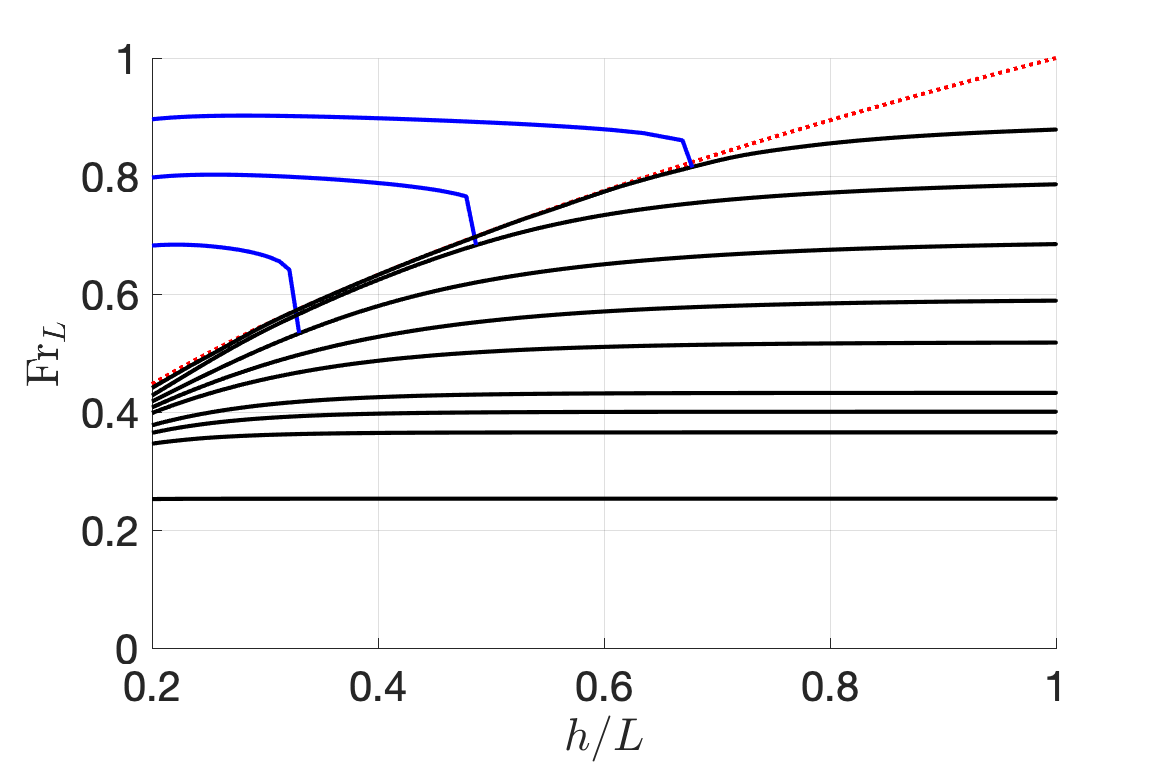}};
\draw[line width=1,red,->] (0.5,-1) -- (0,1.7);
\node at (-3,2) {(a)};
\node[black] at (-1,1.8) {Increasing $F/mg$};
\node[red] at (1.4,1.7) { $\mathrm{Fr}_L=\sqrt{h/L}$};
\node[red] at (1.4,2) { Fr${_h=1}$};
\node at (6,0) {\includegraphics[width=0.45\textwidth]{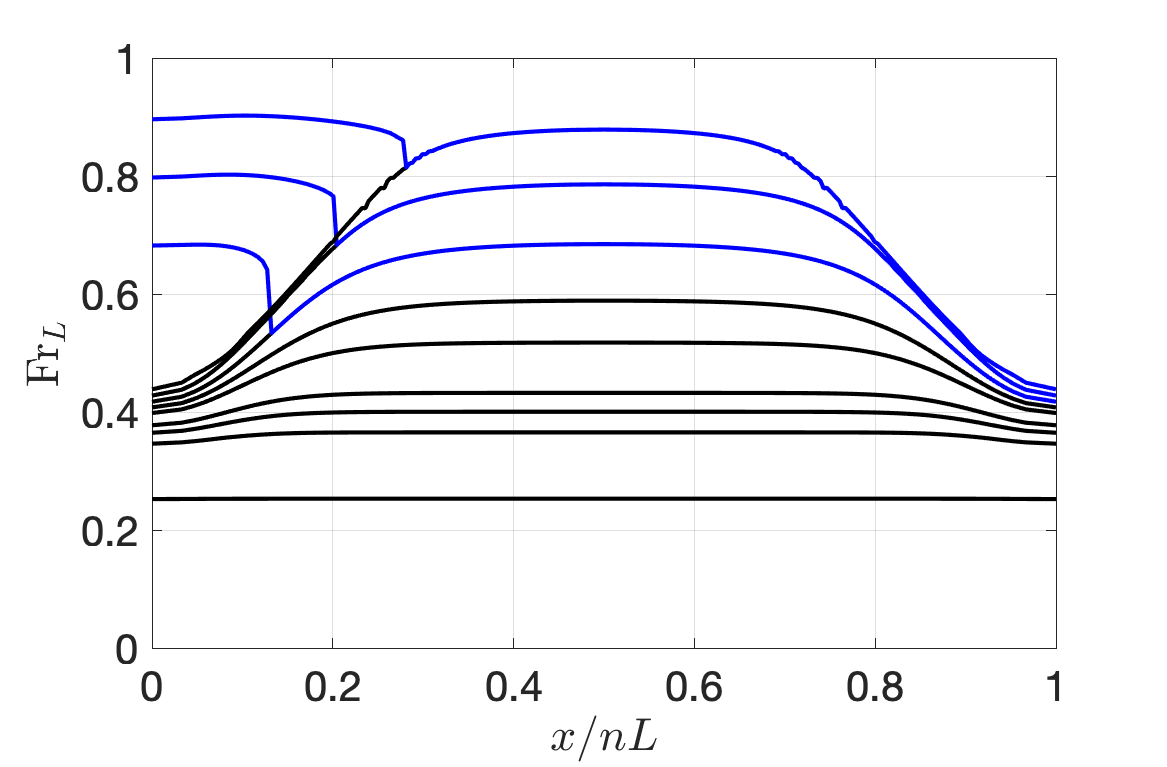}};
\node at (3,-3.25) {\includegraphics[width=0.45\textwidth]{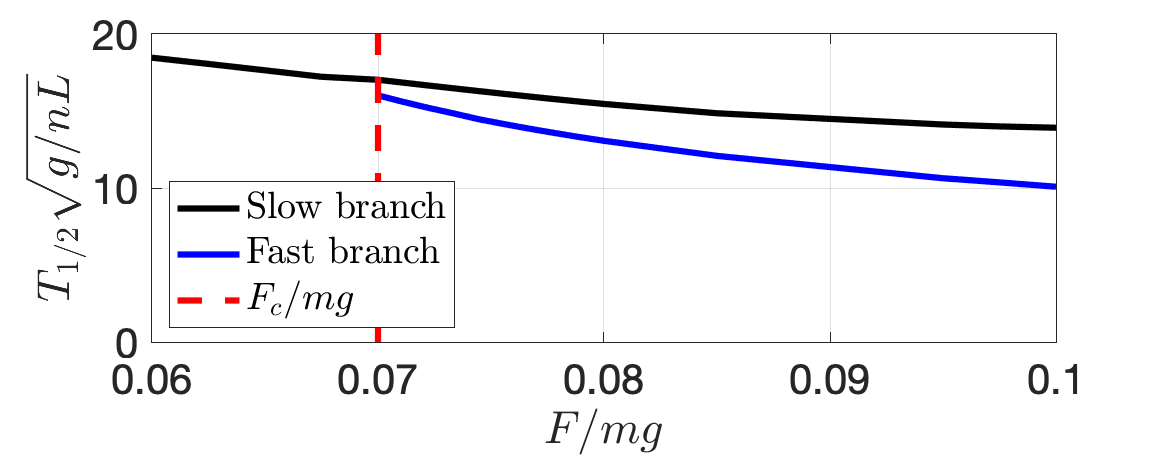}};
\node at (3,1.5) {(b)};
\node at (3,-1.3) {(c)};
\draw[line width=1,red,->] (6,-1) -- (6,1.55);
\node at (6,1.8) {Increasing $F/mg$};
\end{tikzpicture}
\caption{Phase plane analysis of the case study of a boat moving in a lake with sinusoidal depth \eqref{sindepth}, where the rate of change of depth is sufficiently small that the quasi-steady approximation is appropriate. Trajectories for different values of the forcing term $F/mg$ are illustrated in both the depth-Froude plane (a) and the distance-Froude plane (b). (c) Times to complete the first half of the race $0\leq x\leq nL/2$ for the different solution branches.   \label{quasipic}}
\end{figure}
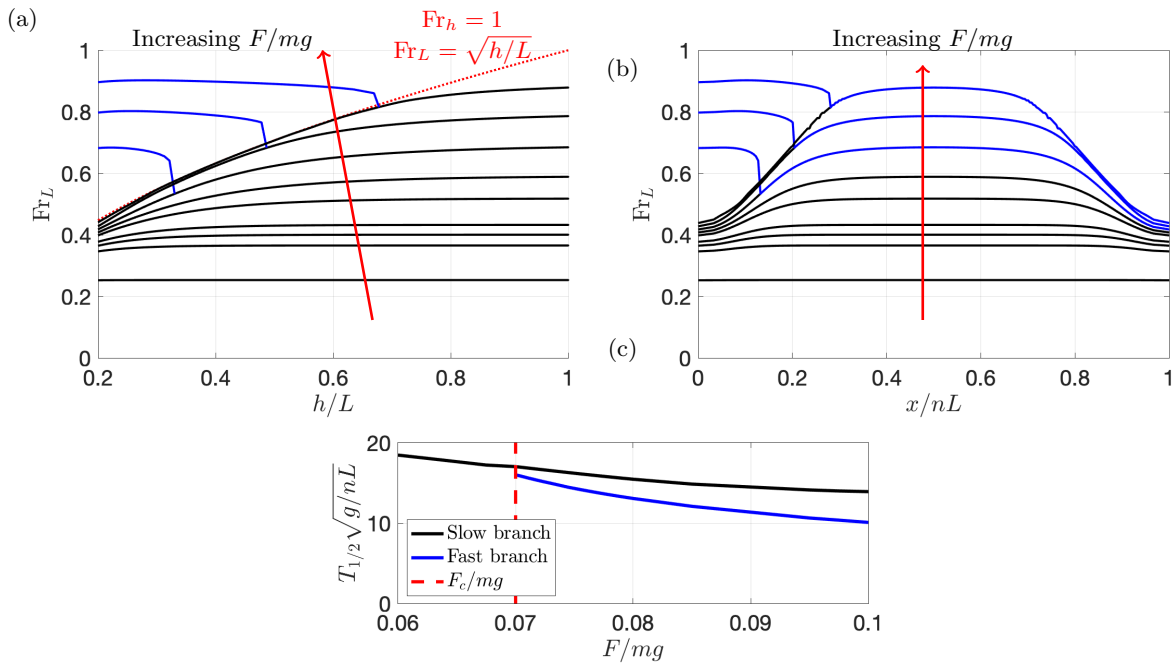

We consider a sequence of forcing magnitudes $F_i/mg$ between $0.005$ and $0.11$, and investigate the quasi-steady dynamics in each of these cases. Such forcing values cover the range relevant for rowing races, where typical forces are around $6-8\%$ of the total boat weight \citep{boucher2018thin}. At $x=0$, where the depth is at its minimum value, two possible stable solutions are available for all forcing values in the range $0.07<F/mg<0.11$. Therefore, for each forcing value $i$ we choose initial conditions for the system \eqref{sys1}-\eqref{sys2} as one of these two branches
\beq
    x_{i,j}(0)=0,\quad u_{i,j}(0)=u^*_{i,j},\quad j=1,2,
\eeq
where the subscript $j$ indicates the branch number, and the starting velocity $u^*_{i,j}$ is one of the two stable solutions to the equilibrium equation
\beq
R_d(0,u^*_i)=F_i.
\eeq
Then, we solve the system \eqref{sys1}-\eqref{sys2} for $0<t<T$, where the final time is determined by the condition 
$x(T)=nL.$
We plot various solutions in Fig. \ref{quasipic}, where we indicate different stable branches with different colours. Fig. \ref{quasipic}(a) shows the solutions on the phase plane generated by $h(t)/L$ and $\mathrm{Fr}_L(t)$. 
Fig. \ref{quasipic}(b) shows the solutions $\mathrm{Fr}_L(t)$ plotted against ${x}(t)/nL$.

The behaviour is remarkably different depending on the forcing magnitude. If the forcing is less than a critical value $F_c/mg=0.07$ (see Fig. \ref{bifurcs}(a)), then only one solution branch exists for all values of $x$. The drag is largest in the shallow part and smallest in the deep part, with perfectly symmetric motion about $x=nL/2$. On the other hand, for $F>F_c$ the catastrophic emergence of multiple solution branches and a hysteresis pattern is observed. One branch is a very slow solution, forced beneath a drag barrier at Fr$_L=\sqrt{h/L}$ (or Fr$_h=1$), and behaves much like the solutions observed for small forcing. The other stable branch of the solution is much faster - in fact, even faster than the equilibrium solution at infinite depth. 
However, on the return journey both branches collapse to the same trajectory due to the fact that the fold bifurcation is topologically disconnected.
Nevertheless, boats which stay on this stable branch have a significantly better race time than the former ones.

We plot the race times $T_{1/2}$ for the first half of the race $0\leq x\leq nL/2$, for various values of the forcing in \ref{quasipic}(c). We give the times in terms of a dimensional scaling for the total race $\sqrt{nL/g}$, and we illustrate where the two stable branches split at the critical forcing. 
The time for the second half of the race is ignored, since both branches converge to the same path and hence this is less interesting. Well above the critical forcing magnitude $F_c/mg$, race times between branches can differ by as much as $30\%$. This critical value is important because force values $F>F_c$ present an opportunity to beat records in racing applications.

\section{Discussion and perspectives}

We have investigated the quasi-steady dynamics of a body moving near the surface of a region of water with slowly varying depth. This work sheds light on previously unstudied bifurcation behaviour that emerges in shallow water. In particular, the two key parameters are the depth $h/L$ and the forcing $F/mg$, which can produce as many as three different solution branches (two stable and one unstable) with potentially large differences in speeds. We showed how this can produce hysteresis routes, and can have significant impact on the time to complete a race in a course with varying depth. In particular, there exists a critical forcing value $F_c$ above which there is an opportunity to accelerate as one enters a shallow patch of water, or be quashed by a rising drag barrier, depending on one's particular branch of the solution.

There are several interesting perspectives following on from this work. Firstly, in boat sports, a comprehensive study of the effect of race bathymetry on performance for different athletes could be investigated and compared with our results.
It would also be interesting to study the interactions between multiple bodies moving in proximity, and how this affects the results presented here. 


{Secondly, as mentioned in the introduction, the non-monotone drag-velocity relationship observed here has also been observed in sharply stratified \textit{dead water}, and Ekman even predicted possible hysteresis effects \cite{ekman1904dead,mercier2011resurrecting}. It would be interesting to extend this work to study the nonlinear dynamics involved when a boat moves in water with a stratification that varies spatially. Such situations are relevant when navigating near a glacial run-off, where the location and strength of the water density stratification may vary as one moves.}
\\

\begin{acknowledgments}
We acknowledge the support from Ecole Polytechnique for the research program Sciences 2024. We also thank JP Boucher and R Labb\'{e} for their help in building the tow-tank facility. 
\end{acknowledgments}

\bibliographystyle{apsrev4-1}
\bibliography{bibfile.bib}

\begin{thebibliography}{29}%
\makeatletter
\providecommand \@ifxundefined [1]{%
 \@ifx{#1\undefined}
}%
\providecommand \@ifnum [1]{%
 \ifnum #1\expandafter \@firstoftwo
 \else \expandafter \@secondoftwo
 \fi
}%
\providecommand \@ifx [1]{%
 \ifx #1\expandafter \@firstoftwo
 \else \expandafter \@secondoftwo
 \fi
}%
\providecommand \natexlab [1]{#1}%
\providecommand \enquote  [1]{``#1''}%
\providecommand \bibnamefont  [1]{#1}%
\providecommand \bibfnamefont [1]{#1}%
\providecommand \citenamefont [1]{#1}%
\providecommand \href@noop [0]{\@secondoftwo}%
\providecommand \href [0]{\begingroup \@sanitize@url \@href}%
\providecommand \@href[1]{\@@startlink{#1}\@@href}%
\providecommand \@@href[1]{\endgroup#1\@@endlink}%
\providecommand \@sanitize@url [0]{\catcode `\\12\catcode `\$12\catcode
  `\&12\catcode `\#12\catcode `\^12\catcode `\_12\catcode `\%12\relax}%
\providecommand \@@startlink[1]{}%
\providecommand \@@endlink[0]{}%
\providecommand \url  [0]{\begingroup\@sanitize@url \@url }%
\providecommand \@url [1]{\endgroup\@href {#1}{\urlprefix }}%
\providecommand \urlprefix  [0]{URL }%
\providecommand \Eprint [0]{\href }%
\providecommand \doibase [0]{http://dx.doi.org/}%
\providecommand \selectlanguage [0]{\@gobble}%
\providecommand \bibinfo  [0]{\@secondoftwo}%
\providecommand \bibfield  [0]{\@secondoftwo}%
\providecommand \translation [1]{[#1]}%
\providecommand \BibitemOpen [0]{}%
\providecommand \bibitemStop [0]{}%
\providecommand \bibitemNoStop [0]{.\EOS\space}%
\providecommand \EOS [0]{\spacefactor3000\relax}%
\providecommand \BibitemShut  [1]{\csname bibitem#1\endcsname}%
\let\auto@bib@innerbib\@empty
\bibitem [{\citenamefont {Boucher}\ \emph {et~al.}(2018)\citenamefont
  {Boucher}, \citenamefont {Labb{\'e}}, \citenamefont {Clanet},\ and\
  \citenamefont {Benzaquen}}]{boucher2018thin}%
  \BibitemOpen
  \bibfield  {author} {\bibinfo {author} {\bibfnamefont {J.}~\bibnamefont
  {Boucher}}, \bibinfo {author} {\bibfnamefont {R.}~\bibnamefont {Labb{\'e}}},
  \bibinfo {author} {\bibfnamefont {C.}~\bibnamefont {Clanet}}, \ and\ \bibinfo
  {author} {\bibfnamefont {M.}~\bibnamefont {Benzaquen}},\ }\href@noop {}
  {\bibfield  {journal} {\bibinfo  {journal} {Phys. Rev. Fluids}\ }\textbf
  {\bibinfo {volume} {3}},\ \bibinfo {pages} {074802} (\bibinfo {year}
  {2018})}\BibitemShut {NoStop}%
\bibitem [{\citenamefont {Fourdrinoy}\ \emph {et~al.}(2019)\citenamefont
  {Fourdrinoy}, \citenamefont {Caplier}, \citenamefont {Devaux}, \citenamefont
  {Rousseaux}, \citenamefont {Gianni}, \citenamefont {Zacharias}, \citenamefont
  {Jouteur}, \citenamefont {Martin}, \citenamefont {Dambrine}, \citenamefont
  {Petcu} \emph {et~al.}}]{fourdrinoy2019naval}%
  \BibitemOpen
  \bibfield  {author} {\bibinfo {author} {\bibfnamefont {J.}~\bibnamefont
  {Fourdrinoy}}, \bibinfo {author} {\bibfnamefont {C.}~\bibnamefont {Caplier}},
  \bibinfo {author} {\bibfnamefont {Y.}~\bibnamefont {Devaux}}, \bibinfo
  {author} {\bibfnamefont {G.}~\bibnamefont {Rousseaux}}, \bibinfo {author}
  {\bibfnamefont {A.}~\bibnamefont {Gianni}}, \bibinfo {author} {\bibfnamefont
  {I.}~\bibnamefont {Zacharias}}, \bibinfo {author} {\bibfnamefont
  {I.}~\bibnamefont {Jouteur}}, \bibinfo {author} {\bibfnamefont
  {P.}~\bibnamefont {Martin}}, \bibinfo {author} {\bibfnamefont
  {J.}~\bibnamefont {Dambrine}}, \bibinfo {author} {\bibfnamefont
  {M.}~\bibnamefont {Petcu}},  \emph {et~al.},\ }in\ \href@noop {} {\emph
  {\bibinfo {booktitle} {5th MASHCON}}}\ (\bibinfo  {publisher} {Flanders
  Hydraulics Research; Maritime Technology Division, Ghent University},\
  \bibinfo {year} {2019})\ pp.\ \bibinfo {pages} {104--133}\BibitemShut
  {NoStop}%
\bibitem [{\citenamefont {Havelock}(1922)}]{havelock1922effect}%
  \BibitemOpen
  \bibfield  {author} {\bibinfo {author} {\bibfnamefont {T.}~\bibnamefont
  {Havelock}},\ }\href@noop {} {\bibfield  {journal} {\bibinfo  {journal}
  {Proc. R. Soc. Lond. A}\ }\textbf {\bibinfo {volume} {100}},\ \bibinfo
  {pages} {499} (\bibinfo {year} {1922})}\BibitemShut {NoStop}%
\bibitem [{\citenamefont {Michell}(1898)}]{michell1898xi}%
  \BibitemOpen
  \bibfield  {author} {\bibinfo {author} {\bibfnamefont {J.}~\bibnamefont
  {Michell}},\ }\href@noop {} {\bibfield  {journal} {\bibinfo  {journal} {The
  London, Edinburgh, and Dublin Philosophical Magazine and Journal of Science}\
  }\textbf {\bibinfo {volume} {45}},\ \bibinfo {pages} {106} (\bibinfo {year}
  {1898})}\BibitemShut {NoStop}%
\bibitem [{\citenamefont {Videler}(2012)}]{videler2012fish}%
  \BibitemOpen
  \bibfield  {author} {\bibinfo {author} {\bibfnamefont {J.}~\bibnamefont
  {Videler}},\ }\href@noop {} {\emph {\bibinfo {title} {Fish swimming}}}\
  (\bibinfo  {publisher} {Springer Science \& Business Media},\ \bibinfo {year}
  {2012})\BibitemShut {NoStop}%
\bibitem [{\citenamefont {Tuck}(1989)}]{tuck1989wave}%
  \BibitemOpen
  \bibfield  {author} {\bibinfo {author} {\bibfnamefont {E.}~\bibnamefont
  {Tuck}},\ }\href@noop {} {\bibfield  {journal} {\bibinfo  {journal} {The
  ANZIAM Journal}\ }\textbf {\bibinfo {volume} {30}},\ \bibinfo {pages} {365}
  (\bibinfo {year} {1989})}\BibitemShut {NoStop}%
\bibitem [{\citenamefont {Hofman}\ and\ \citenamefont
  {Kozarski}(2000)}]{hofman2000shallow}%
  \BibitemOpen
  \bibfield  {author} {\bibinfo {author} {\bibfnamefont {M.}~\bibnamefont
  {Hofman}}\ and\ \bibinfo {author} {\bibfnamefont {V.}~\bibnamefont
  {Kozarski}},\ }\href@noop {} {\bibfield  {journal} {\bibinfo  {journal}
  {International shipbuilding progress}\ }\textbf {\bibinfo {volume} {47}},\
  \bibinfo {pages} {61} (\bibinfo {year} {2000})}\BibitemShut {NoStop}%
\bibitem [{\citenamefont {Sretensky}(1937)}]{sretensky1937theoretical}%
  \BibitemOpen
  \bibfield  {author} {\bibinfo {author} {\bibfnamefont {L.}~\bibnamefont
  {Sretensky}},\ }\href@noop {} {\bibfield  {journal} {\bibinfo  {journal}
  {Joukovsky Cent Inst Rep}\ }\textbf {\bibinfo {volume} {319}},\ \bibinfo
  {pages} {1} (\bibinfo {year} {1937})}\BibitemShut {NoStop}%
\bibitem [{\citenamefont {Weinblum}(1950)}]{weinblum1950analysis}%
  \BibitemOpen
  \bibfield  {author} {\bibinfo {author} {\bibfnamefont {G.}~\bibnamefont
  {Weinblum}},\ }\href@noop {} {\emph {\bibinfo {title} {Analysis of wave
  resistance}}}\ (\bibinfo  {publisher} {Navy Department, the David W. Taylor
  Model Basin},\ \bibinfo {year} {1950})\BibitemShut {NoStop}%
\bibitem [{\citenamefont {Doctors}\ and\ \citenamefont
  {Dagan}(1980)}]{doctors1980comparison}%
  \BibitemOpen
  \bibfield  {author} {\bibinfo {author} {\bibfnamefont {L.}~\bibnamefont
  {Doctors}}\ and\ \bibinfo {author} {\bibfnamefont {G.}~\bibnamefont
  {Dagan}},\ }\href@noop {} {\bibfield  {journal} {\bibinfo  {journal} {J.
  Fluid. Mech.}\ }\textbf {\bibinfo {volume} {98}},\ \bibinfo {pages} {647}
  (\bibinfo {year} {1980})}\BibitemShut {NoStop}%
\bibitem [{\citenamefont {Doctors}(1975)}]{doctors1975experimental}%
  \BibitemOpen
  \bibfield  {author} {\bibinfo {author} {\bibfnamefont {L.}~\bibnamefont
  {Doctors}},\ }\href@noop {} {\bibfield  {journal} {\bibinfo  {journal} {J.
  Fluid. Mech.}\ }\textbf {\bibinfo {volume} {72}},\ \bibinfo {pages} {513}
  (\bibinfo {year} {1975})}\BibitemShut {NoStop}%
\bibitem [{\citenamefont {Li}\ and\ \citenamefont
  {Ellingsen}(2016)}]{li2016ship}%
  \BibitemOpen
  \bibfield  {author} {\bibinfo {author} {\bibfnamefont {Y.}~\bibnamefont
  {Li}}\ and\ \bibinfo {author} {\bibfnamefont {S.}~\bibnamefont {Ellingsen}},\
  }\href@noop {} {\bibfield  {journal} {\bibinfo  {journal} {J. Fluid. Mech.}\
  }\textbf {\bibinfo {volume} {791}},\ \bibinfo {pages} {539} (\bibinfo {year}
  {2016})}\BibitemShut {NoStop}%
\bibitem [{\citenamefont {W{{e}}do{\l}owski}\ and\ \citenamefont
  {Napi{\'o}rkowski}(2013)}]{wkedolowski2013capillary}%
  \BibitemOpen
  \bibfield  {author} {\bibinfo {author} {\bibfnamefont {K.}~\bibnamefont
  {W{{e}}do{\l}owski}}\ and\ \bibinfo {author} {\bibfnamefont {M.}~\bibnamefont
  {Napi{\'o}rkowski}},\ }\href@noop {} {\bibfield  {journal} {\bibinfo
  {journal} {Phys. Rev. E}\ }\textbf {\bibinfo {volume} {88}},\ \bibinfo
  {pages} {043014} (\bibinfo {year} {2013})}\BibitemShut {NoStop}%
\bibitem [{\citenamefont {Millward}\ \emph {et~al.}(1986)\citenamefont
  {Millward}, \citenamefont {Bevan} \emph {et~al.}}]{millward1986effect}%
  \BibitemOpen
  \bibfield  {author} {\bibinfo {author} {\bibfnamefont {A.}~\bibnamefont
  {Millward}}, \bibinfo {author} {\bibfnamefont {M.}~\bibnamefont {Bevan}},
  \emph {et~al.},\ }\href@noop {} {\bibfield  {journal} {\bibinfo  {journal}
  {Journal of ship research}\ }\textbf {\bibinfo {volume} {30}},\ \bibinfo
  {pages} {85} (\bibinfo {year} {1986})}\BibitemShut {NoStop}%
\bibitem [{\citenamefont {Hofman}(2006)}]{hofman2006prediction}%
  \BibitemOpen
  \bibfield  {author} {\bibinfo {author} {\bibfnamefont {M.}~\bibnamefont
  {Hofman}},\ }\href@noop {} {\emph {\bibinfo {title} {Prediction of wave
  making resistance of fast ships in shallow water and computer program
  {S}hallowres}}},\ \bibinfo {type} {Tech. Rep.}\ (\bibinfo {year}
  {2006})\BibitemShut {NoStop}%
\bibitem [{\citenamefont {Ekman}(1904)}]{ekman1904dead}%
  \BibitemOpen
  \bibfield  {author} {\bibinfo {author} {\bibfnamefont {V.}~\bibnamefont
  {Ekman}},\ }\href@noop {} {\bibfield  {journal} {\bibinfo  {journal}
  {Norwegian North Polar Expedition, 1893-1896}\ ,\ \bibinfo {pages} {1}}
  (\bibinfo {year} {1904})}\BibitemShut {NoStop}%
\bibitem [{\citenamefont {Mercier}\ \emph {et~al.}(2011)\citenamefont
  {Mercier}, \citenamefont {Vasseur},\ and\ \citenamefont
  {Dauxois}}]{mercier2011resurrecting}%
  \BibitemOpen
  \bibfield  {author} {\bibinfo {author} {\bibfnamefont {M.}~\bibnamefont
  {Mercier}}, \bibinfo {author} {\bibfnamefont {R.}~\bibnamefont {Vasseur}}, \
  and\ \bibinfo {author} {\bibfnamefont {T.}~\bibnamefont {Dauxois}},\
  }\href@noop {} {\bibfield  {journal} {\bibinfo  {journal} {arXiv preprint
  arXiv:1103.0903}\ } (\bibinfo {year} {2011})}\BibitemShut {NoStop}%
\bibitem [{\citenamefont {Esmaeilpour}\ \emph {et~al.}(2018)\citenamefont
  {Esmaeilpour}, \citenamefont {Ezequiel~Martin},\ and\ \citenamefont
  {Carrica}}]{esmaeilpour2018computational}%
  \BibitemOpen
  \bibfield  {author} {\bibinfo {author} {\bibfnamefont {M.}~\bibnamefont
  {Esmaeilpour}}, \bibinfo {author} {\bibfnamefont {J.}~\bibnamefont
  {Ezequiel~Martin}}, \ and\ \bibinfo {author} {\bibfnamefont {P.~M.}\
  \bibnamefont {Carrica}},\ }\href@noop {} {\bibfield  {journal} {\bibinfo
  {journal} {Journal of Fluids Engineering}\ }\textbf {\bibinfo {volume} {140}}
  (\bibinfo {year} {2018})}\BibitemShut {NoStop}%
\bibitem [{\citenamefont {Medjdoub}\ \emph {et~al.}(2020)\citenamefont
  {Medjdoub}, \citenamefont {J{\'a}nosi},\ and\ \citenamefont
  {Vincze}}]{medjdoub2020laboratory}%
  \BibitemOpen
  \bibfield  {author} {\bibinfo {author} {\bibfnamefont {K.}~\bibnamefont
  {Medjdoub}}, \bibinfo {author} {\bibfnamefont {I.~M.}\ \bibnamefont
  {J{\'a}nosi}}, \ and\ \bibinfo {author} {\bibfnamefont {M.}~\bibnamefont
  {Vincze}},\ }\href@noop {} {\bibfield  {journal} {\bibinfo  {journal}
  {Experiments in Fluids}\ }\textbf {\bibinfo {volume} {61}},\ \bibinfo {pages}
  {6} (\bibinfo {year} {2020})}\BibitemShut {NoStop}%
\bibitem [{\citenamefont {Tuck}(1987)}]{tuck1987wave}%
  \BibitemOpen
  \bibfield  {author} {\bibinfo {author} {\bibfnamefont {E.~O.}\ \bibnamefont
  {Tuck}},\ }\href@noop {} {\bibfield  {journal} {\bibinfo  {journal} {Applied
  Mathematics Report T8701}\ } (\bibinfo {year} {1987})}\BibitemShut {NoStop}%
\bibitem [{\citenamefont {Kracht}(1978)}]{kracht1978design}%
  \BibitemOpen
  \bibfield  {author} {\bibinfo {author} {\bibfnamefont {A.~M.}\ \bibnamefont
  {Kracht}},\ }\href@noop {} {\bibfield  {journal} {\bibinfo  {journal} {SNAME
  Transactions}\ }\textbf {\bibinfo {volume} {86}},\ \bibinfo {pages} {197}
  (\bibinfo {year} {1978})}\BibitemShut {NoStop}%
\bibitem [{\citenamefont {Gourlay}(2003)}]{gourlay2003ship}%
  \BibitemOpen
  \bibfield  {author} {\bibinfo {author} {\bibfnamefont {T.}~\bibnamefont
  {Gourlay}},\ }\href@noop {} {\bibfield  {journal} {\bibinfo  {journal} {Int.
  J. Marit. Eng}\ }\textbf {\bibinfo {volume} {145}},\ \bibinfo {pages} {1}
  (\bibinfo {year} {2003})}\BibitemShut {NoStop}%
\bibitem [{\citenamefont {Benham}\ \emph {et~al.}(2019)\citenamefont {Benham},
  \citenamefont {Boucher}, \citenamefont {Labb{\'e}}, \citenamefont
  {Benzaquen},\ and\ \citenamefont {Clanet}}]{benham2019wave}%
  \BibitemOpen
  \bibfield  {author} {\bibinfo {author} {\bibfnamefont {G.}~\bibnamefont
  {Benham}}, \bibinfo {author} {\bibfnamefont {J.}~\bibnamefont {Boucher}},
  \bibinfo {author} {\bibfnamefont {R.}~\bibnamefont {Labb{\'e}}}, \bibinfo
  {author} {\bibfnamefont {M.}~\bibnamefont {Benzaquen}}, \ and\ \bibinfo
  {author} {\bibfnamefont {C.}~\bibnamefont {Clanet}},\ }\href@noop {}
  {\bibfield  {journal} {\bibinfo  {journal} {J. Fluid. Mech.}\ }\textbf
  {\bibinfo {volume} {878}},\ \bibinfo {pages} {147} (\bibinfo {year}
  {2019})}\BibitemShut {NoStop}%
\bibitem [{\citenamefont {Newman}(2018)}]{newman2018marine}%
  \BibitemOpen
  \bibfield  {author} {\bibinfo {author} {\bibfnamefont {J.}~\bibnamefont
  {Newman}},\ }\href@noop {} {\emph {\bibinfo {title} {Marine hydrodynamics}}}\
  (\bibinfo  {publisher} {MIT press},\ \bibinfo {year} {2018})\BibitemShut
  {NoStop}%
\bibitem [{\citenamefont {Rabaud}\ and\ \citenamefont
  {Moisy}(2014)}]{rabaud2014narrow}%
  \BibitemOpen
  \bibfield  {author} {\bibinfo {author} {\bibfnamefont {M.}~\bibnamefont
  {Rabaud}}\ and\ \bibinfo {author} {\bibfnamefont {F.}~\bibnamefont {Moisy}},\
  }\href@noop {} {\bibfield  {journal} {\bibinfo  {journal} {Ocean
  Engineering}\ }\textbf {\bibinfo {volume} {90}},\ \bibinfo {pages} {34}
  (\bibinfo {year} {2014})}\BibitemShut {NoStop}%
\bibitem [{\citenamefont {Gertler}(1954)}]{gertler1954reanalysis}%
  \BibitemOpen
  \bibfield  {author} {\bibinfo {author} {\bibfnamefont {M.}~\bibnamefont
  {Gertler}},\ }\href@noop {} {\emph {\bibinfo {title} {A reanalysis of the
  original test data for the Taylor Standard Series}}},\ \bibinfo {type} {Tech.
  Rep.}\ (\bibinfo  {institution} {David Taylor Model Basin Washington DC},\
  \bibinfo {year} {1954})\BibitemShut {NoStop}%
\bibitem [{\citenamefont {Benzaquen}\ \emph {et~al.}(2014)\citenamefont
  {Benzaquen}, \citenamefont {Darmon},\ and\ \citenamefont
  {Rapha{\"e}l}}]{benzaquen2014wake}%
  \BibitemOpen
  \bibfield  {author} {\bibinfo {author} {\bibfnamefont {M.}~\bibnamefont
  {Benzaquen}}, \bibinfo {author} {\bibfnamefont {A.}~\bibnamefont {Darmon}}, \
  and\ \bibinfo {author} {\bibfnamefont {E.}~\bibnamefont {Rapha{\"e}l}},\
  }\href@noop {} {\bibfield  {journal} {\bibinfo  {journal} {Physics of
  Fluids}\ }\textbf {\bibinfo {volume} {26}},\ \bibinfo {pages} {092106}
  (\bibinfo {year} {2014})}\BibitemShut {NoStop}%
\bibitem [{\citenamefont {Menter}(1994)}]{menter1992improved}%
  \BibitemOpen
  \bibfield  {author} {\bibinfo {author} {\bibfnamefont {F.}~\bibnamefont
  {Menter}},\ }\href@noop {} {\bibfield  {journal} {\bibinfo  {journal} {AIAA
  Journal}\ }\textbf {\bibinfo {volume} {32}},\ \bibinfo {pages} {1598}
  (\bibinfo {year} {1994})}\BibitemShut {NoStop}%
\bibitem [{\citenamefont {Zhu}\ \emph {et~al.}(2015)\citenamefont {Zhu},
  \citenamefont {He}, \citenamefont {Zhang}, \citenamefont {Wu}, \citenamefont
  {Wan}, \citenamefont {Zhu},\ and\ \citenamefont
  {Noblesse}}]{zhu2015farfield}%
  \BibitemOpen
  \bibfield  {author} {\bibinfo {author} {\bibfnamefont {Y.}~\bibnamefont
  {Zhu}}, \bibinfo {author} {\bibfnamefont {J.}~\bibnamefont {He}}, \bibinfo
  {author} {\bibfnamefont {C.}~\bibnamefont {Zhang}}, \bibinfo {author}
  {\bibfnamefont {H.}~\bibnamefont {Wu}}, \bibinfo {author} {\bibfnamefont
  {D.}~\bibnamefont {Wan}}, \bibinfo {author} {\bibfnamefont {R.}~\bibnamefont
  {Zhu}}, \ and\ \bibinfo {author} {\bibfnamefont {F.}~\bibnamefont
  {Noblesse}},\ }\href@noop {} {\bibfield  {journal} {\bibinfo  {journal}
  {European Journal of Mechanics-B/Fluids}\ }\textbf {\bibinfo {volume} {49}},\
  \bibinfo {pages} {226} (\bibinfo {year} {2015})}\BibitemShut {NoStop}%
\end{thebibliography}%

\end{document}